\documentstyle[12pt,twoside,fleqn,epsfig]{article}

\begin{document}

\title{Assessment of the $GW$ approximation using Hubbard chains}
\author{Thomas J Pollehn\dag, Arno Schindlmayr\dag\thanks{Corresponding
author. E-mail address: as10031@phy.cam.ac.uk} \ and R W Godby\ddag\\[1ex]
\hspace{-10.15pt}\small \dag\ Cavendish Laboratory, University of
Cambridge, Madingley Road, Cambridge CB3 0HE, UK\\
\small \ddag\ Department of Physics, University of York, Heslington, York
YO1 5DD, UK}
\date{Received 19 May 1997}
\maketitle

\begin{abstract}
We investigate the performance of the $GW$ approximation by comparison to
exact results for small model systems. The role of the chemical potentials
in Dyson's equation as well as the consequences of numerical resonance
broadening are examined, and we show how a proper treatment can improve
computational implementations of many-body perturbation theory in general.
$GW$ and exchange-only calculations are performed over a wide range of
fractional band fillings and correlation strengths. We thus identify the
physical situations where these schemes are applicable.
\end{abstract}


\section{Introduction}

Materials with strong electronic correlation are of considerable interest
in solid state science, but their computational treatment is notoriously
difficult: diagonalizing the corresponding Hamiltonians is not feasible for
large systems, and the strong correlation causes mean-field methods to
break down. The collective dynamics of such systems can in principle be
described exactly by many-body perturbation theory, however. In this
framework, all exchange and correlation effects are absorbed into the
self-energy operator $\Sigma$, which may be thought of as a non-local,
energy-dependent potential. In this paper, we investigate a class of
self-energies based on Hedin's $GW$ approximation \cite{Hed65}. Their
diagrammatic representation, which neglects explicit vertex corrections, is
reminiscent of the Fock exchange potential, but the Coulomb interaction
includes dynamic screening. Our first aim is to examine whether numerical
improvements can be achieved by including further correlation
effects in the underlying propagators without changing the diagrammatic
form of the self-energy. To assess the performance of these schemes, we
compare the calculated spectra with exact results for small model systems
that can still be solved by numerical diagonalization techniques. A similar
study was recently reported for a two-dimensional Hubbard cluster
\cite{Ver95}; here we extend that work by considering further variants as
well as a larger variety of systems, most importantly a much wider range of
band fillings. Our second aim is to optimize the practical implementation.
To this end, we investigate the treatment of the chemical potentials in
Dyson's equation and the consequences of resonance broadening in the course
of numerical manipulations. Both points are all too often ignored, but may
have significant impact on calculated spectra.

The central quantity of interest is the one-particle Green's function $G$,
whose imaginary part is directly linked to the spectral function
$A = \pi^{-1} | \,\mbox{Im}\, G |$. While many authors adopt a momentum and
energy representation $G({\bf k},\omega)$ that follows naturally from the
band theory of extended systems, we will always consider
$G_{{\bf RR}'}(\omega)$ in real space, which is more appropriate for finite
clusters. Furthermore, this representation has the advantage of showing the
entire excitation spectrum in each diagonal element of the Green's
function. While this is not always desirable if one wants to concentrate on
the evolution of particular quasiparticles, it allows us to judge the
performance of any self-energy approximation on the basis of a single
matrix element. However, we also calculated $G$ in reciprocal space for
corresponding translationally invariant systems to confirm our
identification of various spectral features with either quasiparticles or
satellites from a particular excitation.

The order of this paper is as follows. Section 2 introduces the model
Hamiltonian. In section 3 we describe our procedure for obtaining the exact
Green's function. In section 4 we review the $GW$ approximation and address
details of the practical implementation. Section 5 lists the approximation
variants we consider and gives numerical results. Finally, section 6
contains our conclusions.

\section{Model description}

The Hubbard model \cite{Hub63} is the classic example of a Hamiltonian that
describes strong, short-range electron-electron interaction. It is
sufficiently simple to be diagonalized exactly for small cluster sizes
using standard numerical techniques, yet its physical behaviour is
non-trivial and reflects many properties of real materials. The model
variant we employ is a finite chain of $M$ ions with open boundary
conditions. Each lattice site contains one orbital that can accommodate up
to two electrons with opposite spin. Doubly occupied orbitals are penalized
by a repulsive on-site interaction $U$, while the hopping of transient
electrons between neighbouring sites yields an energy gain $-t$. The full
Hamiltonian is
\begin{equation} \label{eq:genericHubbard}
{\cal H} = - t \sum_{\langle{\bf R},{\bf R}'\rangle,\sigma} c^\dagger_{{\bf
R}\sigma} c_{{\bf R}'\sigma} + U \sum_{\bf R}\hat{n}_{{\bf R}\uparrow}
\hat{n}_{{\bf R}\downarrow} + \sum_{{\bf R},\sigma} V_{\bf R}
\hat{n}_{{\bf R}\sigma}
\end{equation}
where $c^\dagger_{{\bf R}\sigma}$,$c_{{\bf R}\sigma}$ are the creation and
annihilation operators for an electron at site ${\bf R}$ with spin
$\sigma$, $\hat{n}_{{\bf R}\sigma} \equiv c^\dagger_{{\bf R}\sigma}
c_{{\bf R}\sigma}$ is the particle number operator, and $\langle {\bf R},
{\bf R}' \rangle$ indicates a sum over nearest neighbours only. We choose
the energy norm by setting $t = 1$. The Hamiltonian further
contains a local potential $V_{\bf R}$ that will later serve as a
mean-field approximation for exchange and correlation. We denote the total
electron number by $N$.

The properties of the Hubbard model have been thoroughly investigated. In
particular, the one-dimensional case can be solved analytically using the
Bethe ansatz \cite{Lie68} and, in the limit of infinite chain length, is
known to yield a Luttinger-liquid ground state. The corresponding Green's
function describes a gapless spectrum of bosonic collective modes involving
charge and spin degrees of freedom \cite{Med92}. For finite $M$, however,
the renormalized quasiparticle weight factors remain non-zero as long as
the Coulomb integral $U$ does not exceed a critical value \cite{Qin95},
which behaves asymptotically like $1/M$. In this parameter range the model
exhibits Fermi-liquid behaviour. The resolution required to differentiate
convincingly between Fermi liquids and Luttinger liquids is in fact near
infinitesimal on the energy scale we consider, and the Lorentzian
broadening of resonances essentially wipes out features on a genuinely
small scale that are of primary concern in the distinction between the two.
Even so, we have confirmed that all systems we study in this paper are
comfortably within the Fermi-liquid regime, so that the same perturbation
methods as for higher dimensions can be applied.

\begin{figure}[t]
\epsfysize=6.5cm \centerline{\epsfbox{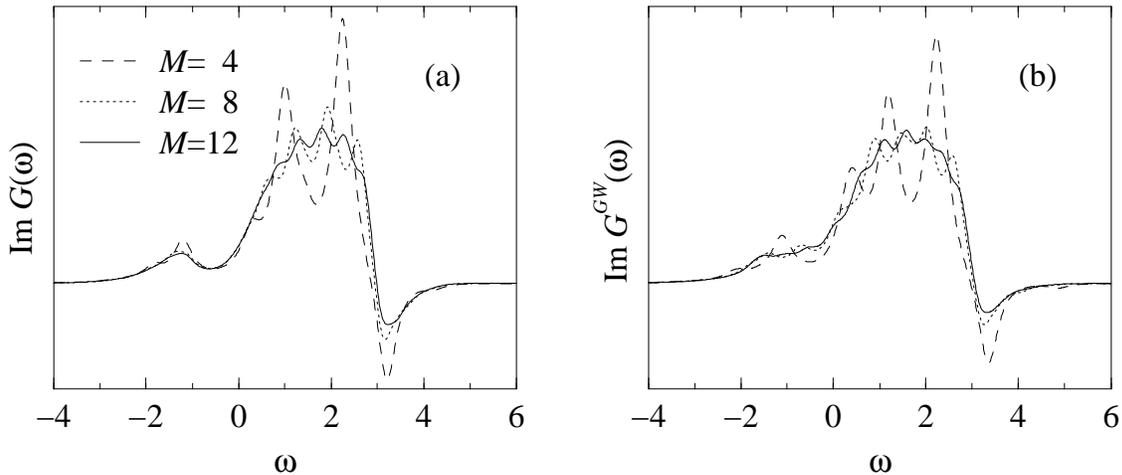}}
\caption{\rm Exact Green's function (a) and $GW$ approximation (b) for
varying chain length $M$ with constant 75\% band filling and $U = 2$. The
broad quasiparticle and satellite spectral features are insensitive to $M$,
indicating that within the Fermi-liquid limitations of the model it is
possible to generalize the results reported here. \label{fig:length_func}}
\end{figure}

As we are working with small model Hamiltonians, it is essential to
consider the possible sensitivity of our results to the parameters in
(\ref{eq:genericHubbard}). The system size is a particularly important
aspect. We have calculated the exact Green's function and a correponding
$GW$ approximation for varying chain length while keeping $U = 2$ and the
fractional band filling $N / (2M) = 75\%$ constant. In figure
\ref{fig:length_func}, as in all later graphs, we show the matrix element
$\mbox{Im}\, G_{1,1}(\omega)$ in arbitrary units and align the chemical
potentials to faciliate comparison. The number of peaks in the spectral
function grows with the chain length as expected. However, it is also
evident that the {\em qualitative\/} appearance of the graphs changes
little: the broad quasiparticle and satellite peaks are insensitive to $M$.
This becomes even clearer when the integral $\int_{-\infty}^{\omega}
\mbox{Im}\, G(\omega') \,{\rm d}\omega'$, which averages over oscillations
on a small scale, is considered. We have thus demonstrated that the results
reported in the following sections are relatively insensitive to the chain
length and so retain significance beyond the particular model geometry,
although of course the formal extrapolation to $M \to \infty$ cannot be
made because of the eventual transition to a Luttinger liquid.

\section{Exact numerical solution}

The exact one-particle Green's function at zero temperature is defined as
\begin{equation} \label{eq:g_exact_in_t}
G_{{\bf RR}'}(t-t') = -{\rm i} \langle N | {\cal T} \{ c_{{\bf R}\sigma}(t)
c^\dagger_{{\bf R}'\sigma}(t') \} | N \rangle
\end{equation}
where $| N \rangle$ is the ground state of the interacting $N$-electron
system, ${\cal T}$ is Wick's time-ordering operator, and
$c_{{\bf R}\sigma}(t) \equiv \exp({\rm i}{\cal H}t) c_{{\bf R}\sigma}
\exp(-{\rm i}{\cal H}t)$ denotes the time-dependent wave field operator in
the Heisenberg picture. We have suppressed the spin index in $G$ because
the Green's function is diagonal and degenerate in $\sigma$. It is
convenient to Fourier transform (\ref{eq:g_exact_in_t}) to the energy
domain and rewrite the Green's function in the form
\begin{equation} \label{eq:g_exact_in_e}
G_{{\bf RR}'}(\omega) = \langle N | c_{{\bf R}\sigma} \frac{1}{\omega
- {\cal H}^+ + E_N} c^\dagger_{{\bf R}'\sigma} | N \rangle + \langle N |
c^\dagger_{{\bf R}'\sigma} \frac{1}{\omega + {\cal H}^- - E_N}
c_{{\bf R}\sigma} |N \rangle .
\end{equation}
Here $E_N$ is the ground-state energy corresponding to $| N \rangle$ and
${\cal H}^\pm$ denotes the Hamiltonian matrix for $N \pm 1$ electrons.

The main computational difficulty is that the number of basis vectors of
the many-body problem grows exponentially with the system size, because it
quantifies the $(2M!)$\ $/ [(2M-N)!N!]$ possibilities of distributing $N$
electrons onto $M$ two-fold spin-degenerate orbitals. For a ten-site chain
at half filling, the largest model we consider, this implies a basis size
of 184,756 (although greater chain lengths are feasible if the band filling
is very small or very large). However, less than 0.01\% of the elements of
${\cal H}$ are non-zero, so that sparse-matrix techniques may be used to
obtain $| N \rangle$ and $E_N$.

The diagonal elements $G_{\bf RR}$, which enter the calculation of the
electron density and other quantities, may be calculated without full
matrix inversion by tridiagonalizing $\omega \mp {\cal H}^{\pm} \pm E_N$
using the recursion method \cite{Hay80} and starting with the vector
$c^\dagger_{{\bf R}'\sigma} | N \rangle$ or $c_{{\bf R}\sigma} | N
\rangle$. For non-diagonal elements a block recursion must be performed.
Neither an inversion nor a complete diagonalization would be feasible in
terms of computer memory: for the example $M = N = 10$ mentioned above, 210
GB are required to store the eigenvectors of ${\cal H}^\pm$ and 254 GB for
the inverse of $\omega \mp {\cal H}^\pm \pm E_N$, although this could be
somewhat reduced by exploiting symmetry relations.

Once the recursion coefficients, i.e., the diagonal elements $a_n$ and the
off-diagonal elements $b_n^2$ of the tridiagonal matrix, are determined
iteratively up to a suitable recursion depth $D$, the elements of the
Green's function are obtained from
\begin{equation} \label{eq:g_exact_recursion}
G_{\bf RR}(\omega) = \frac{1}{\displaystyle \omega - a_0 -
\frac{b_1^2}{\displaystyle \omega - a_1 - \cdots -
\frac{b_D^2}{\displaystyle \omega - a_D}}} .
\end{equation}
Even with a basis size of 184,756 about 400 recursions are sufficient,
since the number of actual spectral features is small compared to the basis
size. Ideally a single recursive level for each additional peak in the
spectrum would be necessary, and indeed we require only a few recursions
per feature to achieve full convergence in practice. The quasiparticles and
collective excitations of the $N$-electron system are determined by the
eigenvalues of ${\cal H}^\pm$ and feature as simple poles in the Green's
function. For numerical convenience, we broaden these sharp resonances into
Lorentzians by offsetting the singularities from the real energy axis by a
distance $\delta$. This procedure does not imply a finite lifetime of the
excited states.

\section{The $GW$ approximation}

The $GW$ approximation constitutes a diagrammatic expansion of the
self-energy that neglects explicit vertex corrections. However, it includes
dynamic screening of the Coulomb interaction and is thus capable of
describing certain correlation effects. Originally the $GW$ self-energy was
derived as a first order iterative solution of Hedin's coupled equations
for the propagators of the interacting many-electron system starting from
Hartree theory \cite{Hed65}. In this section we briefly review the
formalism and subsequently address crucial details concerning the
computational implementation.

\subsection{The self-energy in the $GW$ approximation}

Starting from a mean-field Hamiltonian that may contain a suitable
effective potential $V$, we obtain a zeroth order Green's function
\begin{equation}
G^0_{{\bf RR}'}(\omega) = \sum_s \frac{\langle {\bf R} | \psi_s \rangle
\langle \psi_s | {\bf R'} \rangle}{\omega - \epsilon_s + {\rm i}
\,\mbox{sgn} (\epsilon_s - \mu_0) \delta}
\end{equation}
in terms of the one-particle eigenstates $| \psi_s \rangle$ and
corresponding energy eigenvalues $\epsilon_s$. The symbol $\mu_0$ denotes
the chemical potential. Conventionally, the random phase approximation
\begin{equation} \label{eq:bubbleP}
P^{\rm RPA}_{{\bf RR}'}(\tau) = -2 {\rm i} G^0_{{\bf RR}'}(\tau)
G^0_{{\bf R}'{\bf R}}(-\tau)
\end{equation}
for the irreducible polarization propagator is employed, with $\tau \equiv
t - t'$ and a factor 2 for spin summation. The dynamically screened
interaction is then obtained from
\begin{equation} \label{HowToGetW}
W^{\rm RPA}(\omega) = U \left[ 1 - P^{\rm RPA}(\omega) U \right]^{-1}
\end{equation}
in matrix notation. Finally, the self-energy is given by the expression
\begin{equation} \label{GWselfenergy}
\Sigma^{GW}_{{\bf RR}'}(\tau) = {\rm i} G^0_{{\bf RR}'}(\tau)
W^{\rm RPA}_{{\bf RR}'}(\tau + \eta)
\end{equation}
to which the $GW$ approximation owes its name. $\eta$ denotes a positive
infinitesimal. The self-energy may be inserted into Dyson's equation to
yield the improved Green's function
\begin{equation} \label{DysonEq}
G^{GW}(\omega) = \left[ 1 - G^0(\omega) \left( V^{\rm H} +
\Sigma^{GW}(\omega) - V \right) \right]^{-1} G^0(\omega)
\end{equation}
where $V^{\rm H}_{\bf R} = U \langle \hat{n}_{{\bf R}\uparrow} +
\hat{n}_{{\bf R}\downarrow} \rangle$ indicates the Hartree potential.
During the calculation we use Fast Fourier Transforms to change between the
time and energy domains as appropriate in order to avoid costly numerical
convolutions.

\subsection{Alignment of the chemical potentials}

Dyson's equation (\ref{DysonEq}) combines the equation of motion of the
interacting with that of the corresponding non-interacting system. The
self-energy specifies the deviation of the quasiparticle states from the
bare electrons and holes upon adiabatic introduction of the Coulomb
potential. It is thus an equilibrium quantity that should really be
calculated self-consistently, i.e., the dressed Green's function obtained
from Dyson's equation is reinserted into the self-energy until convergence
has been achieved. This procedure is so computationally demanding, however,
as to make it unfeasible for large-scale {\it ab initio} calculations.
Furthermore, while the random phase approximation $P^{\rm RPA}[G^0]$ by
construction gives the proper response function of time-dependent Hartree
theory, the same expression evaluated with a self-consistent Green's
function ceases to yield a physically meaningful propagator due to the
neglect of appropriate vertex corrections. As a consequence the spectrum
becomes broad and structureless \cite{Bar96}. On the other hand, the
self-energy inherits the chemical potential of the underlying Green's
function, and a mismatch with that of the dressed propagator may result in
wrong time-ordering.

A possible solution is to use a zeroth order $G^0$ to evaluate the
self-energy but shift it in such a way as to align its chemical potential
with that of the dressed Green's function. This limited degree of
self-consistency, originally suggested by Hedin \cite{Hed65}, suffices to
ensure the correct time-ordering while leaving the response function
unchanged and thus physically meaningful. From the Fourier transform of
(\ref{GWselfenergy}), we see that shifting $G^0$ by an amount
$\tilde{\omega}$ on the energy axis translates into an identical shift of
the self-energy
\begin{equation}
\Sigma^{GW}_{{\bf RR}'}(\omega-\tilde{\omega}) = \frac{\rm i}{2\pi}
\int\!\! G^0_{{\bf RR}'}(\omega-\tilde{\omega}-\omega') W^{\rm RPA}_{{\bf
RR}'}(\omega') {\rm e}^{{\rm i} \eta \omega'} \,{\rm d}\omega'
\end{equation}
where the contour is closed about the upper half-plane. According to
Dyson's equation, the chemical potential of the dressed Green's function
becomes $\mu = \mu_0 + \langle V^{\rm H} + \Sigma^{GW}(\mu-\tilde{\omega})
- V \rangle$, where the matrix element is to be taken with the highest
occupied quasiparticle orbital. In practice we use the corresponding
$| \psi_s \rangle$ of the non-interacting system. The shift is determined
so that the chemical potential coincides exactly with that of the relocated
zeroth order Green's function, hence $\mu = \mu_0 + \tilde{\omega}$.
Inserting this relation into the previous equation yields the explicit
solution
\begin{equation}
\tilde{\omega} = \langle V^{\rm H} + \Sigma^{GW}(\mu_0) - V \rangle .
\end{equation}

\begin{figure}[t]
\epsfysize=7cm \centerline{\epsfbox{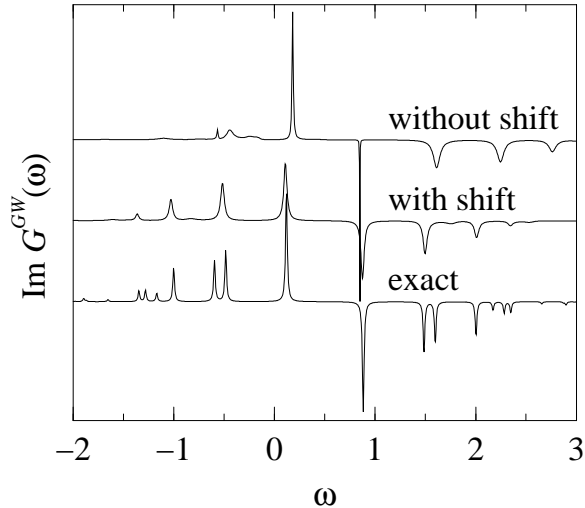}}
\caption{\rm The self-consistency shift $\tilde{\omega}$, which aligns the
chemical potential of the zeroth order $G^0$ with that of the dressed
Green's function derived from it, improves the spectral features
substantially even for a very weak interaction of $U = 1$. The exact
spectrum is shown for comparison. \label{fig:align_ferm}}
\end{figure}

Despite its early suggestion, this self-consistency shift is often ignored
in {\it ab initio} band structure calculations, where its impact is
normally small. It may substantially improve the more sensitive satellite
spectrum, however. As an example we consider an eight-site chain at
half-filling with $U = 1$. In figure \ref{fig:align_ferm} we display
results obtained with and without $\tilde{\omega}$. The exact spectrum is
shown for comparison. While the shift has little effect on the
quasiparticle peaks, the improved description of the satellites is evident
even for such a weak interaction. The results follow our previous
demonstration that to a large extent $\tilde{\omega}$ also restores
particle number conservation, which is generally violated in
non-self-consistent many-body theory \cite{Sch97}. All subsequent
calculations incorporate $\tilde{\omega}$.

\subsection{Infinitesimal peak broadening}

\begin{figure}[t]
\epsfysize=7cm \centerline{\epsfbox{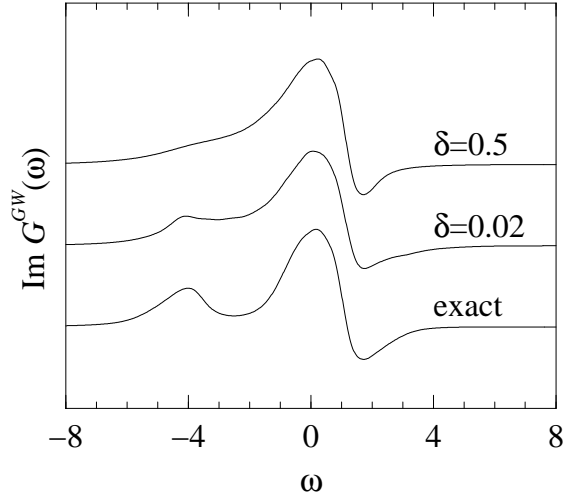}}
\caption{\rm Comparison of spectra calculated with $\delta = 0.5$ and
$\delta = 0.02$, where the latter was broadened afterwards. The smaller
Lorentzian width incurs a higher computational cost but yields more
pronounced spectral features. \label{fig:no_large_delta}}
\end{figure}

The $GW$ approximation constitutes a diagrammatic expansion of the true
self-energy to first order in the screened interaction. The underlying
equations (\ref{eq:bubbleP}) to (\ref{DysonEq}) may be solved in the time
as well as in the energy domain. While we employ Fast Fourier Transforms to
switch between the two as appropriate, most implementations work
exclusively in the latter, where all operations are either multiplicative
or feasible but computationally costly convolutions. It is frequently
ignored, however, that the relations are only strictly valid in the limit
$\delta \to 0$, because convolutions with a finite displacement of the
singularities from the energy axis will mix the real and imaginary parts of
the propagators. As a consequence, weak features such as satellites tend to
smear out and may even become undetectable in the calculated spectrum. As
an example, figure \ref{fig:no_large_delta} contrasts the spectral
functions from two calculations with $\delta = 0.5$ and $\delta = 0.02$,
where the resonances in the latter were suitably broadened afterwards for
the purpose of comparison. We also show the exact spectrum. The model
specifications are $M = 10$, $N = 14$ and medium interaction $U = 4$. The
features in the second curve are clearly more pronounced. In particular,
the reproduction of the satellite at $-4$ is superior. The downside of
using small values for $\delta$ is an increased number of sampling points,
because the energy resolution in a numerical treatment must necessarily
exceed the characteristic peak width. In the following we always choose
$\delta$ as small as computationally possible and only broaden the final
spectra to achieve a Lorentzian width of 0.5 for visualization.

\section{Comparison of different approximation variants}

In this section we investigate in detail several variants of the $GW$
scheme by direct comparison with exact results. The main criterion we apply
to judge approximations is their ability to reproduce the overall shape of
the true spectral function, i.e., the position and weight of the
quasiparticle excitations as well as their satellites.

\subsection{Different initial Hamiltonians}

\begin{figure}[t]
\epsfysize=7cm \centerline{\epsfbox{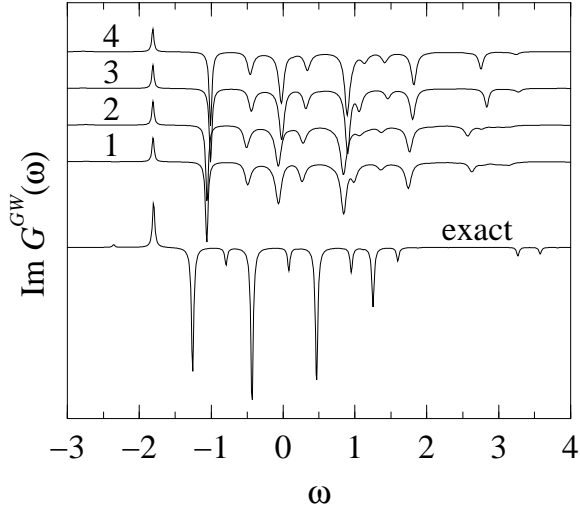}}
\caption{\rm Using a renormalized Green's function or screening in the $GW$
self-energy without vertex corrections fails to improve the spectrum.
The numbered curves indicate the combinations (1) $G^0 W^{\rm RPA}[G^0]$,
(2) $G W^{\rm RPA}[G^0]$, (3) $G^0 W^{\rm RPA}[G]$ and (4) $G W^{\rm
RPA}[G]$ for two electrons on a twelve-site chain with $U = 2$.
\label{fig:g_exact}}
\end{figure}

The framework outlined in section 4 allows for considerable freedom in the
choice of the initial mean-field Hamiltonian. We here consider the
following options in order of increasing complexity.
\begin{enumerate}
\item Hedin's original iterative derivation suggests that the Hartree
approximation $V = V^{\rm H}$ with self-consistently determined site
occupation numbers should be used, although this is rarely done in
practice.
\item The most common choice in {\it ab initio} calculations is to start
with a self-consistent exchange-correlation potential from
density-functional (DF) theory, which yields the same charge density as the
interacting system. We simulate this procedure by numerically determining a
potential $V = V^{\rm H} + V^{\rm xc}$ that reproduces the occupation
numbers $\langle \hat{n}_{{\bf R}\sigma} \rangle$ of the exact solution. We
still refer to this approach as density-functional theory, although it is
really a {\em site occupation function(al) theory} \cite{Sch95}.
\item We can also evaluate the self-energy (\ref{GWselfenergy}) with the
exact Green's function $G$, which is fully renormalized and contains a
background satellite spectrum. While this approach itself is of course not
directly relevant to practical calculations, it serves as an example for
implementations that attempt to include a maximum amount of many-body
features in the initial Green's function. 
\end{enumerate}
It was long regarded as self-evident that including as much information
about exchange and correlation as possible in the initial zeroth order
Hamiltonian will provide an optimal starting point for the iteration.
Since the extra computational cost required to include a Hartree or
local-density mean-field in the Hamiltonian is negligible compared to a
complete $GW$ calculation, this approach possesses great appeal. As our
model is numerically solvable, we are in fact able to use the exact Green's
function as an extreme example of an improved propagator that tries to
incorporate as many correlation effects as possible up to dynamic
renormalization. Besides the conventional random phase approximation
$W^{\rm RPA}[G^0]$ with density-functional theory as a zeroth order
approximation we have evaluated the same diagrams using the exact Green's
function to obtain the more sophisticated $W^{\rm RPA}[G]$, which contains
a rich satellite spectrum. Note, however, that this is not the exact
screening, since vertex corrections in the polarization are still ignored.
In figure \ref{fig:g_exact} we show the exact spectrum together with the
results from the four possible combinations of these dielectric functions
with $G^0$ and $G$ in the self-energy for two electrons on a twelve-site
chain with $U = 2$. Evidently the four curves differ very little, implying
that renormalization does not improve the spectrum, but the computational
expense is considerably higher if the exact Green's function is used either
in the screening or the self-energy.

The rule emerging here is that there is no particular advantage in
sophisticated, renormalized propagators when one works within the $GW$
scheme. Instead these should be {\em consistent\/} with the current level
of iteration, because the self-energy (\ref{GWselfenergy}) neglects
explicit vertex corrections and so does not become exact when evaluated
using the true propagators. In general, overrealistic propagators may even
cause the approximation to deteriorate, because they destroy the balance
that exists between the internal diagrammatic expansion of the Green's
function and screening on the one hand and the vertex function on the other
\cite{Gro95}. We thus restrict ourselves to Hartree and density-functional
theory as zeroth order approximations in the following and present results
for these cases below.

\subsection{Model dielectric functions}

\begin{figure}[p]
\epsfysize=13.5cm \centerline{\epsfbox{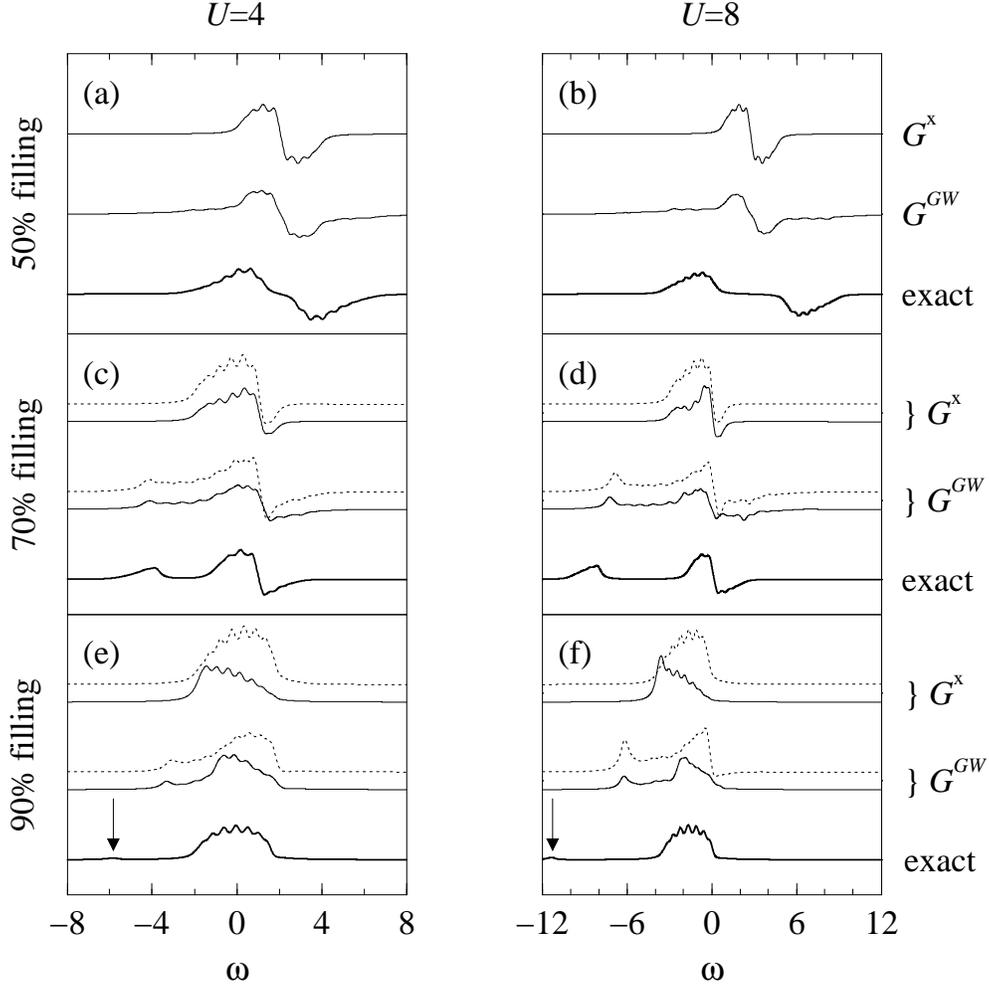}}
\caption{\rm The exact Green's function for a ten-site chain compared to
the $GW$ approximation $G^{GW}$ and the exchange-only $G^{\rm x}$ for
different band fillings $N/(2M)$ and correlation strengths $U$. Where pairs
of curves are shown, the solid line refers to a zeroth order
density-functional and the dotted line to a Hartree Hamiltonian. Some small
satellites are marked by arrows. The $GW$ scheme performs best for
intermediate band filling. The exchange-only scheme yields increasingly
accurate quasiparticles for high band filling and weak correlation but
cannot produce satellites. \label{fig:comparison_mega}}
\end{figure}

As the random phase approximation demands inconvenient numerical
convolutions in the energy domain, practical implementations frequently
deviate from the original $GW$ scheme by employing alternative model
dielectric functions that can be evaluated directly \cite{Eng93}. In order
of increasing complexity, the following options present themselves.
\begin{enumerate}
\item Neglecting screening effects and using the bare Coulomb interaction
is particularly inexpensive, because no intermediate polarization
propagator is required. In this case the self-energy
\begin{equation}
\Sigma^{\rm x}_{{\bf RR}'} = -U \langle \hat{n}_{{\bf R}\sigma} \rangle
\delta_{{\bf RR}'}
\end{equation}
is diagonal and energy-independent. We denote the corresponding Green's
function, which contains electronic exchange but no dynamic correlation, by
$G^{\rm x}$. If performed self-consistently, this approach is identical to
the Hartree--Fock treatment.
\item The screening $W^{\rm RPA}[G^0]$ may be calculated in the random
phase approximation from the zeroth order Green's function in accordance
with the original proposition of the $GW$ approximation, yielding $G^{GW}$.
\item Irrespective of the initial mean-field Hamiltonian, we can employ a
realistic screened interaction that contains more correlation effects.
However, we have already argued earlier that a unilateral expansion of the
screened interaction will not improve the spectrum, and figure
\ref{fig:g_exact} gave a numerical example with $W^{\rm RPA}[G]$ to this
effect. Therefore we will not consider this option further.
\end{enumerate}
In figure \ref{fig:comparison_mega} we show the exact spectral function for
a ten-site chain together with four approximations, namely $G^{GW}$ and
$G^{\rm x}$ evaluated both with Hartree and density-functional theory as
zeroth order Hamiltonians. We increase the band filling from 50\% in two
steps to 90\% and consider the situation of medium ($U = 4$) as well as
strong ($U = 8$) correlation. Due to the particle-hole symmetry of the
Hubbard model, the spectrum for band fillings below 50\% can be obtained by
inflection. For reference, we quote the effective potential parameters
$V_{\bf R}$ for a selected system with 70\% band filling and $U = 4$ in
table \ref{table:g0values}.

\begin{table}[p]
\caption{\rm Effective potential parameters in Hartree and
density-functional (DF) theory for ten sites with 70\% band filling and
$U = 4$. The chain is symmetric about its centre. \label{table:g0values}}
\begin{center}
\begin{tabular}{|l|rrrrr|}
\hline
Site index & 1;10 & 2;9$\;$ & 3;8$\;$ & 4;7$\;$ & 5;6$\;$ \\
\hline
\hline
Hartree & 0.26 & $-0.02$ &   0.06  &   0.10  &   0.06  \\
DF      & 0.26 & $-0.33$ & $-0.08$ & $-0.06$ & $-0.15$ \\
\hline
\end{tabular}
\end{center}
\end{table}

Irrespective of the correlation strength, the occupation numbers are
necessarily uniform at half filling due to particle-hole symmetry. Hence in
this case the dressed Green's functions derived from Hartree and
density-functional theory as starting points coincide except for a constant
shift. The exchange $\Sigma^{\rm x}$ is also uniform and causes just
another constant shift, while the $GW$ approximation improves slightly but
still fails to reproduce the true spectrum satisfactorily, especially for
large $U$. For instance, the energy gap in figure
\ref{fig:comparison_mega}(b) is crucially underestimated. On the plus side,
the approximations have the correct symmetry about the chemical potential.
As the band filling increases, the $GW$ approximation at first becomes
better but again deteriorates for very large band filling. The dominant
interaction processes in this limit are described by the $T$-matrix
\cite{Gal58}, which renormalizes the Hartree--Fock potential by including
multiple scattering in the particle-particle channel to all orders. 
$\Sigma^{GW}$ ignores these diagrams and so does not reproduce the spectrum
well: it is outperformed even by the unrenormalized exchange approximation.
The $GW$ approximation works best for intermediate band filling as in
figures \ref{fig:comparison_mega}(c) and (d), where it yields good results
even for strong correlation. In this regime there is also little difference
between Hartree and density-functional theory as starting points, unlike
for higher band filling where the $GW$ approximation breaks down. The
description of quasiparticles is in general superior to that of satellites,
which are more sensitive to the specific form of many-body interaction
processes. The bare exchange approximation naturally works best in
situations where screening effects are negligible, i.e., if the interaction
is weak and/or the band filling is very high, such as in figure
\ref{fig:comparison_mega}(e). The absence of long-range interaction in the
Hubbard model adds to this effect. Hartree theory as a zeroth order
approximation gives consistently better results than density-functional
theory. Being an effective mean field, $\Sigma^{\rm x}$ of course cannot
produce a satellite structure. This is acceptable, however, because
collective excitations carry little spectral weight in the limits of small
$U$ and high fractional band filling.

\section{Conclusions}

We have examined the implementation and performance of the $GW$
approximation by comparing exact and approximate spectra for finite Hubbard
chains that exhibit Fermi-liquid behaviour. The insensitivity with respect
to the chain length makes the results transferable, although further
investigations for higher-dimensional systems will be useful. Focusing on
the computational implementation, we pointed out that the mathematical
structure of many-body perturbation theory requires a shift aligning the
chemical potentials in Dyson's equation as well as a minimization of the
artificial Lorentzian broadening of spectral peaks. In each case model
calculations clearly demonstrated improvements in the $GW$ approximation,
particularly with respect to the satellite spectrum. Next we studied the
performance of the $GW$ and the related bare exchange approximation for
various band fillings and correlation strengths. Our results show that the
former can yield good results for intermediate band filling even if the
correlation is strong, while the latter provides a computationally cheaper
alternative for weak correlation and small or high band filling. The
description of quasiparticles is generally better than that of satellites.
We have also shown evidence that renormalized propagators will not improve
the $GW$ approximation without the inclusion of appropriate vertex
corrections in the self-energy.

\section*{Acknowledgments}

We are grateful to R Haydock and C M M Nex for helpful discussions. This
work was supported by the Royal Society and the European Community
programme Human Capital and Mobility through contract no.\ CHRX-CT93-0337.
T J Pollehn and A Schindlmayr wish to thank the Studienstiftung des
deutschen Volkes for financial support. A Schindlmayr gratefully
acknowledges further support from the Deutscher Akademischer
Austauschdienst under its HSP III scheme, the Gottlieb Daimler- und Karl
Benz-Stiftung, Pembroke College Cambridge, and the Engineering and Physical
Sciences Research Council.


\begin{thebibliography}{99}
\bibitem{Hed65} Hedin L 1965 {\it Phys. Rev.} {\bf 139} A796
\bibitem{Ver95} Verdozzi C, Godby R W and Holloway S 1995 {\it Phys. Rev.
Lett.} {\bf 74} 2327
\bibitem{Hub63} Hubbard J 1963 {\it Proc. R. Soc.} {\bf 276} 238
\bibitem{Lie68} Lieb E H and Wu F Y 1968 {\it Phys. Rev. Lett.} {\bf 20}
1445
\bibitem{Med92} Meden V and Sch\"onhammer K 1992 {\it Phys. Rev.} B {\bf
46} 15 753
\par\item[] Voit J 1993 {\it Phys. Rev.} B {\bf 47} 6740; 1993 {\it J. Phys.:
Condens. Matter} {\bf 5} 8305
\bibitem{Qin95} Qin S, Qian T, Yu L and Su Z 1995 {\it Phys. Rev.} B {\bf
51} 16 594
\bibitem{Hay80} Haydock R 1980 {\it Solid State Physics} vol 35, ed H
Ehrenreich {\it et al} (New York: Academic) p 216
\bibitem{Bar96} von Barth U and Holm B 1996 {\it Phys. Rev.} B {\bf 54}
8411
\bibitem{Sch97} Schindlmayr A 1997 {\it Phys. Rev.} B {\bf 56} 3528,
cond-mat/9709275
\bibitem{Sch95} Schindlmayr A and Godby R W 1995 {\it Phys. Rev.} B
{\bf 51} 10 427, cond-mat/9709266
\par\item[] Sch\"onhammer K, Gunnarsson O and Noack R M 1995 {\it Phys.
Rev.} B {\bf 52} 2504
\bibitem{Gro95} de Groot H J, Bobbert P A and van Haeringen W 1995 {\it
Phys. Rev.} B {\bf 52} 11 000
\par\item[] Shirley E L 1996 {\it Phys. Rev.} B {\bf 54} 7758
\bibitem{Eng93} Engel G E and Farid B 1993 {\it Phys. Rev.} B {\bf 47}
15 931
\bibitem{Gal58} Galitskii V 1958 {\it Sov. Phys.--JETP} {\bf 7} 104
\end{thebibliography}
\end{document}